\documentclass[conference]{IEEEtran}

\usepackage{amsmath}
\usepackage{graphicx}
\usepackage{booktabs}
\usepackage{balance}
\usepackage[hidelinks]{hyperref}

\DeclareMathOperator{\mean}{mean}
\DeclareMathOperator{\logit}{logit}
\DeclareMathOperator{\BS}{BS}
\DeclareMathOperator{\LogLoss}{LogLoss}
\DeclareMathOperator{\RPS}{RPS}
\DeclareMathOperator{\Score}{Score}
\DeclareMathOperator{\RelativeScore}{RelativeScore}

\begin{document}

\title{Time-Varying Favorite--Longshot Bias in Polymarket Weather Markets}

\author{
\IEEEauthorblockN{Shouchen Qiu}
\IEEEauthorblockA{\textit{High School Student}\\
Grade 11, Class 6\\
Northeast Yucai School\\
Shenyang, China}
\and
\IEEEauthorblockN{Huaijun Qiu}
\IEEEauthorblockA{\textit{Department of Computer Science}\\
\textit{Shenyang Aerospace University}\\
Shenyang, China\\
20220071@email.sau.edu.cn}
}

\maketitle

\begin{abstract}
Prediction-market prices combine traders' judgments about future outcomes, but their errors may be concentrated in particular price ranges and change as settlement approaches. Prior cross-category evidence shows that Polymarket weather markets have larger pricing bias than other major categories, without separating low-price longshots from high-price favorites. This study compares the two ranges at different horizons in daily maximum-temperature markets. Event-level probability scores, cubic B-spline recalibration, and Buy-NO tests are used to examine the form and effect of the bias. Longshot overpricing mainly occurs at distant horizons, whereas favorite overpricing becomes much stronger during the final 1--12 hours and is nearly one order of magnitude larger. Event-level scores improve toward the daily temperature peak, while simple recalibration yields no statistically significant gains. The Buy-NO results in the corresponding time windows also show that the main pricing error is concentrated at the favorite end.
\end{abstract}


\section{Introduction}

Prediction markets turn dispersed information into prices. A binary contract priced at 0.30 is often read as a 30\% chance of success; such an interpretation is calibrated when about 30\% of contracts at that price eventually win \cite{wolfers2004}. Beliefs, risk preferences, and trading behavior can still create systematic errors. The favorite--longshot bias is one such error: low-price and high-price outcomes are mispriced in different directions or by different amounts, and the pattern may change as settlement approaches \cite{page2013}.

Weather markets provide a useful case for studying this bias. Cross-category evidence from Polymarket shows that weather contracts have larger calibration errors than other major categories \cite{akey2026}. That result identifies an unusual market category but does not show whether the error comes from low-price longshots, high-price favorites, or different stages before settlement. This distinction matters because the two cases imply different behavior: traders may overvalue cheap outcomes when uncertainty is high, or become too confident in expensive outcomes when the result appears nearly known.

The central question is how favorite and longshot bias changes with time and which side accounts for the main error in weather markets. An average across all prices can hide opposite movements at the two ends, while one observation horizon cannot show when the bias forms. The analysis therefore follows both ranges from 48 hours to 1 hour before the daily temperature peak. Event-level scores then connect local bias to the quality of the full temperature distribution; a B-spline test examines whether the observed nonlinearity can be corrected by a simple smooth function; and Buy-NO tests examine whether overpricing has an economically consistent direction.

The main contributions are:
\begin{itemize}
\item A time-based comparison separates favorite and longshot bias and identifies the horizons at which each error is strongest.
\item Range bias, full-distribution scores, and smooth recalibration connect local mispricing with the overall quality of the temperature probability distribution.
\item Buy-NO tests in the two strongest bias windows compare the economic effects of favorite and longshot overpricing.
\end{itemize}

\section{Related Work}

\subsection{Favorite--Longshot Bias and Time}

Prediction-market accuracy and probability calibration address related but different questions. Wolfers and Zitzewitz describe how markets collect dispersed beliefs, while noting that the probability meaning of a price depends on market conditions \cite{wolfers2004}. Page and Clemen find that calibration changes with both price and time to settlement \cite{page2013}, and Manski shows that prices do not uniquely identify average belief when traders differ in beliefs or risk preferences \cite{manski2006}. Strong forecasting performance therefore does not guarantee calibration in every price range. Berg et al., for example, find that election prediction markets outperform polls over long horizons \cite{berg2008}; their aggregate accuracy result does not remove the need to inspect errors near the probability tails.

The favorite--longshot literature explains why errors often concentrate near the tails. Ali estimates subjective and objective probabilities from more than 20,000 horse races and finds exaggerated low winning probabilities and depressed high winning probabilities \cite{ali1977}. Racetrack and lottery evidence also shows demand for low-probability, high-payoff longshots \cite{thaler1988}. Snowberg and Wolfers distinguish risk preference from probability misperception as competing explanations \cite{snowberg2010}. Ottaviani and S{\o}rensen instead link the direction and size of the bias to the balance between private information and noise \cite{ottaviani2010}. Because this balance changes as information arrives, time to settlement can change the bias itself.

The existing evidence leaves a setting gap. Racetrack work focuses on parimutuel returns, while broad prediction-market studies combine events with different information processes. Daily temperature markets contain mutually exclusive and ordered outcomes, and public weather information changes continuously before the peak. Their two price tails and time paths therefore require joint analysis under one price definition.

\subsection{Domain Differences on Polymarket}

Calibration on Polymarket differs by event category. Akey et al. compare trading outcomes, order types, and price quality and find especially large bias in weather markets \cite{akey2026}. Le studies Polymarket and Kalshi jointly and finds stable domain and time patterns in calibration \cite{le2026}. These results show that a single all-market calibration curve can hide differences across both categories and horizons.

Cardozo and Rivero-Wildemauwe study favorite--longshot returns and price accuracy across Polymarket categories, including Weather, and show that aggregation choices affect the measured bias \cite{cardozo2026}. Our study complements this evidence by examining how longshot and favorite calibration changes over time in daily maximum-temperature markets.

Cross-category evidence identifies weather as an unusual domain but does not locate the error within a temperature event. Aggregating cheap tail buckets, high-price central buckets, and intermediate outcomes can hide opposite movements at the two ends. Pooling horizons can also mix distant longshot beliefs with near-term favorite confidence. Separating price range from time to the daily peak links the category-level result to individual temperature buckets.

\section{Data}

\subsection{Markets and Sample}

Each daily maximum-temperature event covers one city and date and contains mutually exclusive temperature buckets ordered from low to high. A YES share pays one dollar when its temperature bucket wins and zero otherwise. The buckets therefore describe a probability distribution over the possible daily maximum temperature rather than unrelated binary outcomes. The data span December 5, 2025 to July 31, 2026 and cover ten cities. Table~\ref{tab:data} summarizes 2,189 events, 21,697 bucket contracts, and 8,544,732 taker trades.

Data were obtained through the Polymarket API. The analysis code is available on GitHub\footnote{\url{https://github.com/ShouchenQiu/PolyWeatherData}}.

\begin{table}[!t]
\caption{Dataset summary from December 5, 2025 to July 31, 2026. Events are city-date maximum-temperature markets, Contracts are mutually exclusive temperature-bucket contracts, and Trades are taker transactions used to construct the horizon prices.}
\label{tab:data}
\centering
\footnotesize
\begin{tabular}{lrrr}
\toprule
City & Events & Contracts & Trades \\
\midrule
Atlanta & 246 & 2,366 & 685,156 \\
Chicago & 201 & 2,055 & 581,655 \\
London & 249 & 2,395 & 1,505,627 \\
Miami & 201 & 2,055 & 692,431 \\
New York & 246 & 2,362 & 1,222,144 \\
Seattle & 245 & 2,359 & 620,928 \\
Seoul & 248 & 2,380 & 1,227,963 \\
Shanghai & 151 & 1,655 & 852,606 \\
Tokyo & 154 & 1,682 & 547,984 \\
Toronto & 248 & 2,388 & 608,238 \\
\midrule
Total & 2,189 & 21,697 & 8,544,732 \\
\bottomrule
\end{tabular}
\end{table}

\subsection{Event Structure and Time Split}

For an event with $n$ buckets, $y_k\in\{0,1\}$ is the outcome of bucket $k$. Exactly one bucket wins, so $\sum_{k=1}^{n}y_k=1$.
The observed events usually contain 7--11 buckets. Data are split by event date: 1,585 events from December 5 through May 31 form the training set, 294 June events form the validation set, and 310 July events form the test set. This chronological design limits leakage across nearby periods \cite{tashman2000}; model choices are completed before the final time-out-of-sample evaluation \cite{white2000}.

\section{Experiments and Results}

\subsection{Reliability Analysis}

\begin{figure*}[!t]
\centering
\includegraphics[width=0.90\textwidth]{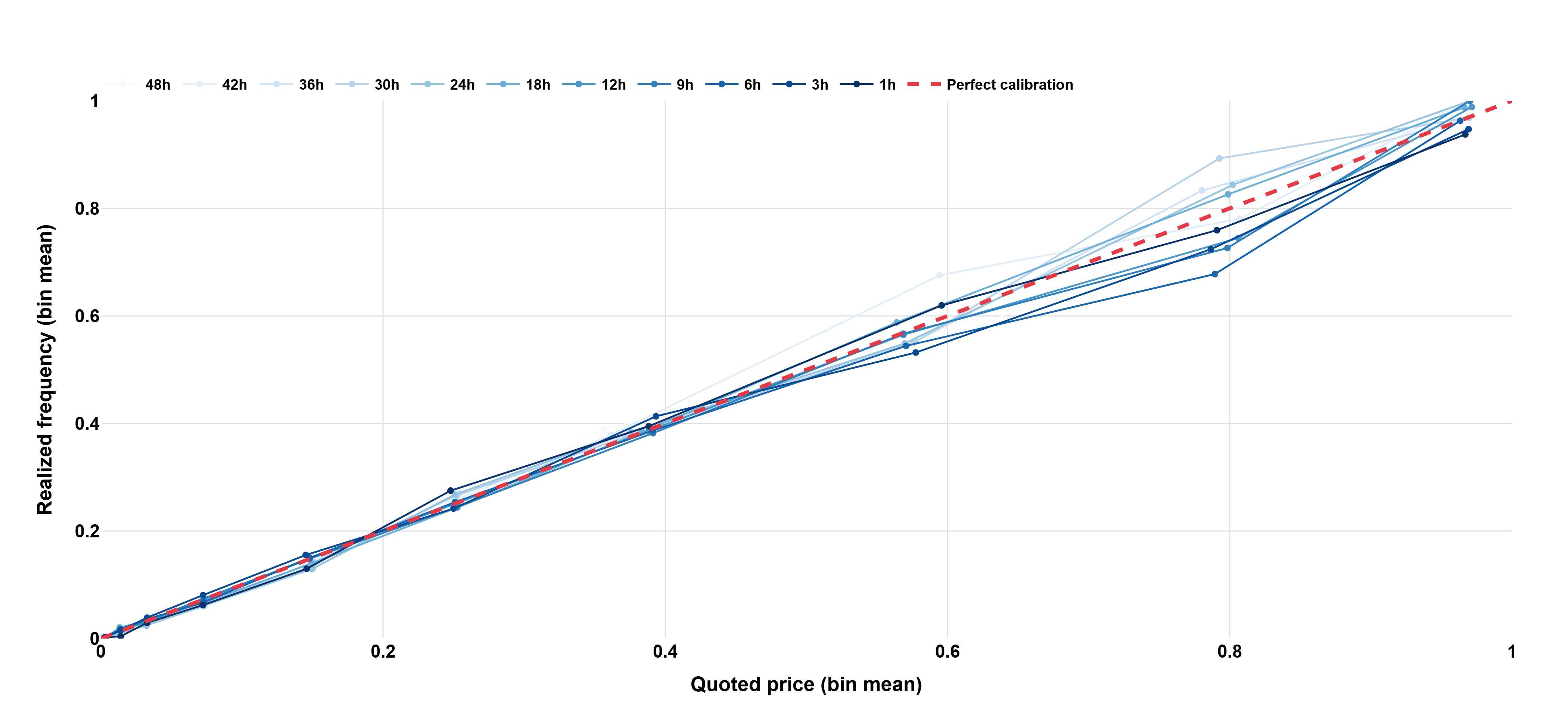}
\caption{Reliability of market prices from 48 to 1 hour before the daily temperature peak. The horizontal axis is the mean quoted price in each bin, and the vertical axis is the realized win rate. Horizons are identified in the legend; darker blue curves indicate shorter horizons. The dashed red diagonal indicates perfect calibration; points below it indicate overpricing. Deviations are largest in the 0.75--0.95 range during the final 1--12 hours.}
\label{fig:reliability}
\end{figure*}

Figures~\ref{fig:reliability} and~\ref{fig:dh} use only the training sample (December 5, 2025--May 31, 2026).

At horizon $h$, the price $p$ of one bucket is the mean of all taker trade prices in the $h\pm0.5$ hour window before the daily temperature peak. Smaller $h$ is closer to the peak. The same definition is used throughout all analyses.

For each price bin, Fig.~\ref{fig:reliability} compares the mean price $\bar p$ with the observed win rate $\bar y$. A point below the diagonal indicates overpricing. Most middle- and low-price bins remain near the diagonal. The favorite end shows a clearer error: during the final 1--12 hours, several bins between 0.75 and 0.95 lie below the diagonal. Although more weather information is available near the peak, traders remain too confident in high-price buckets.

\subsection{Favorite--Longshot Range Bias}

The reliability curves become crowded at the two ends, so range bias is used for direct comparison. For horizon $h$ and price range $[a,b]$,
\begin{equation}
D_h(a,b)=\mean(y-p),\qquad a\leq p\leq b.
\label{eq:dh}
\end{equation}
A negative value means that the market price exceeds the observed win rate. Longshots are defined by $0<p\leq0.10$, a cutoff specified in the research plan. Favorites are defined by $0.75\leq p\leq0.95$. The lower favorite bound isolates clearly high-price outcomes while retaining observations at distant horizons. Horizons are 1, 3, 6, 9, 12, 18, 24, 30, 36, 42, and 48 hours.

\begin{figure*}[!t]
\centering
\includegraphics[width=0.76\textwidth]{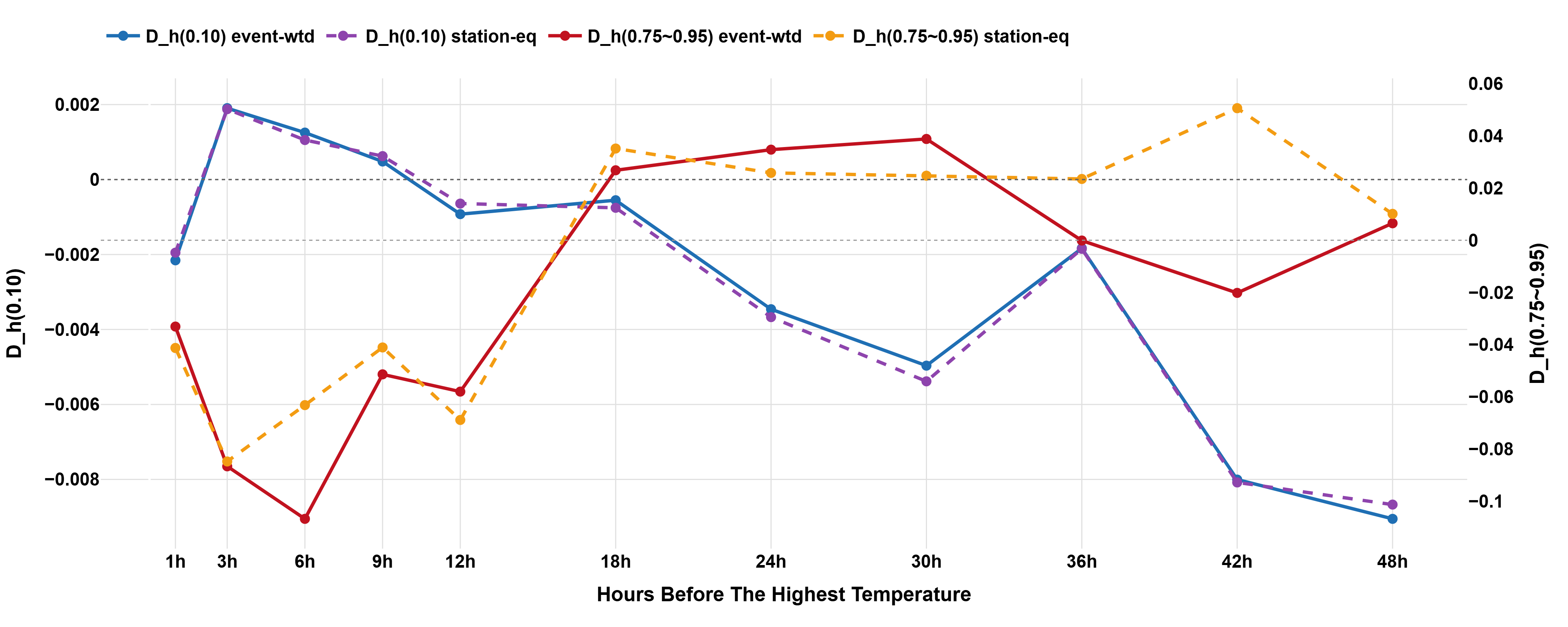}
\caption{Range bias by horizon before the daily temperature peak. Longshots ($0<p\leq0.10$) use the left axis; favorites ($0.75\leq p\leq0.95$) use the right. Solid curves weight events equally; dashed curves also weight the ten cities equally. Negative values indicate overpricing.}
\label{fig:dh}
\end{figure*}

Two weighting schemes control differences in event and city sample sizes. For event weighting (event-wtd), all qualifying buckets are first averaged within each event, and the resulting event values receive equal weight. For station weighting (station-eq), event-wtd bias is first calculated within each city, after which the ten city values receive equal weight. Their close agreement in Fig.~\ref{fig:dh} shows that the pattern is not driven mainly by cities with more events.

Longshot overpricing is concentrated at 30--48 hours. It shrinks as the peak approaches and changes to slight underpricing around 3--9 hours. Traders appear more willing to believe cheap temperature buckets at distant horizons, but reduce that belief once the likely range becomes clearer. Favorites follow the opposite time pattern. Overpricing between 0.75 and 0.95 is strongest during the final 1--12 hours and is nearly one order of magnitude larger than the longshot error (Fig.~\ref{fig:dh}). For example, the event-weighted favorite bias is approximately $-0.10$ at 6 hours, compared with approximately $-0.009$ for longshots at 48 hours. Excess confidence in high-price buckets near the peak is therefore the main source of bias.

\begin{figure}[t]
\centering
\includegraphics[width=\columnwidth,trim=0 48bp 0 0,clip]{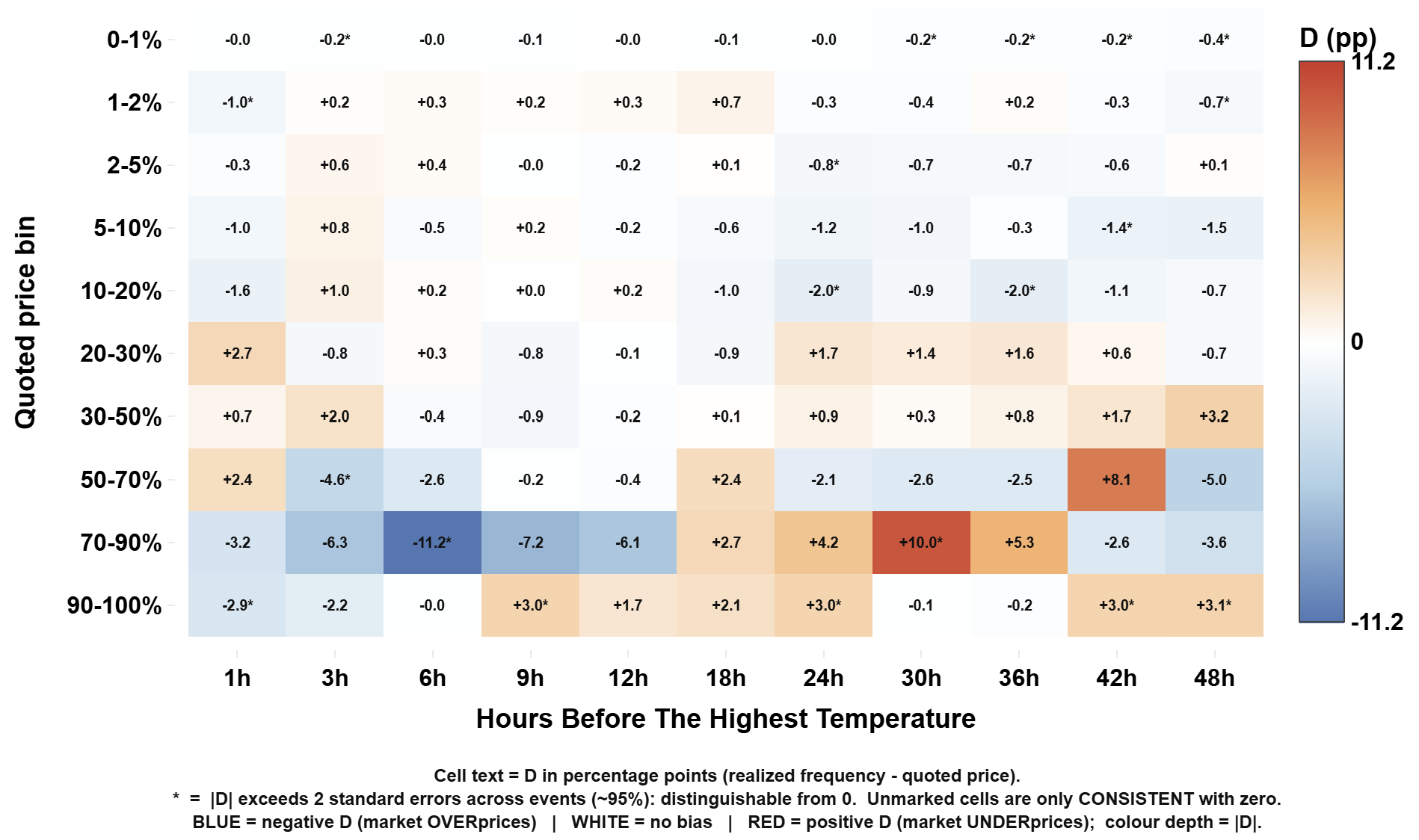}
\caption{Calibration error by price bin and horizon in the training sample. Cell values are realized win rates minus mean prices, in percentage points. Blue indicates overpricing and red underpricing. Asterisks mark errors exceeding two standard errors across events in absolute value.}
\label{fig:heatmap}
\end{figure}

The heatmap shows a similar pattern to Fig.~\ref{fig:dh}: low-price bins exhibit mild overpricing at distant horizons, while stronger near-term overpricing is concentrated in the 70--90\% bin (Fig.~\ref{fig:heatmap}). The latter reaches $-11.2$ percentage points at 6 hours. The heatmap also shows that the errors vary across price bins and horizons rather than shifting uniformly across the distribution.

\subsection{Event-Level Probability Scores}

Range bias uses raw prices to assess the calibration of traded prices, while normalized probabilities are used for event-level scores over all buckets. For event-level scoring, $p_1,\ldots,p_n$ are normalized as
\begin{equation}
S=\sum_{k=1}^{n}p_k,\qquad q_k=\frac{p_k}{S},\qquad \sum_{k=1}^{n}q_k=1.
\label{eq:norm}
\end{equation}
The Brier score measures squared error over the full probability distribution \cite{brier1950}:
\begin{equation}
\BS=\sum_{k=1}^{n}(q_k-y_k)^2.
\label{eq:brier}
\end{equation}
A uniform forecast over $n$ buckets has a Brier score of $1-1/n$, ranging from 0.8571 to 0.9091 for 7--11 buckets.

Log Loss focuses on the probability assigned to the winning bucket and strongly penalizes very low probability on the realized outcome:
\begin{equation}
\LogLoss=-\log(q_{\mathrm{winner}}).
\label{eq:logloss}
\end{equation}
Ranked probability score (RPS) also uses the order of the temperature buckets \cite{epstein1969}. With
\begin{equation}
F_k=\sum_{j=1}^{k}q_j,\qquad O_k=\sum_{j=1}^{k}y_j,
\label{eq:cumulative}
\end{equation}
it is defined as
\begin{equation}
\RPS=\sum_{k=1}^{n-1}(F_k-O_k)^2.
\label{eq:rps}
\end{equation}
RPS gives a larger penalty when probability is placed far from the realized temperature bucket. Together, the three proper scores measure overall allocation, confidence in the winning bucket, and error along the temperature order \cite{gneiting2007}; lower values are better.

\begin{figure}[t]
\centering
\includegraphics[width=\columnwidth]{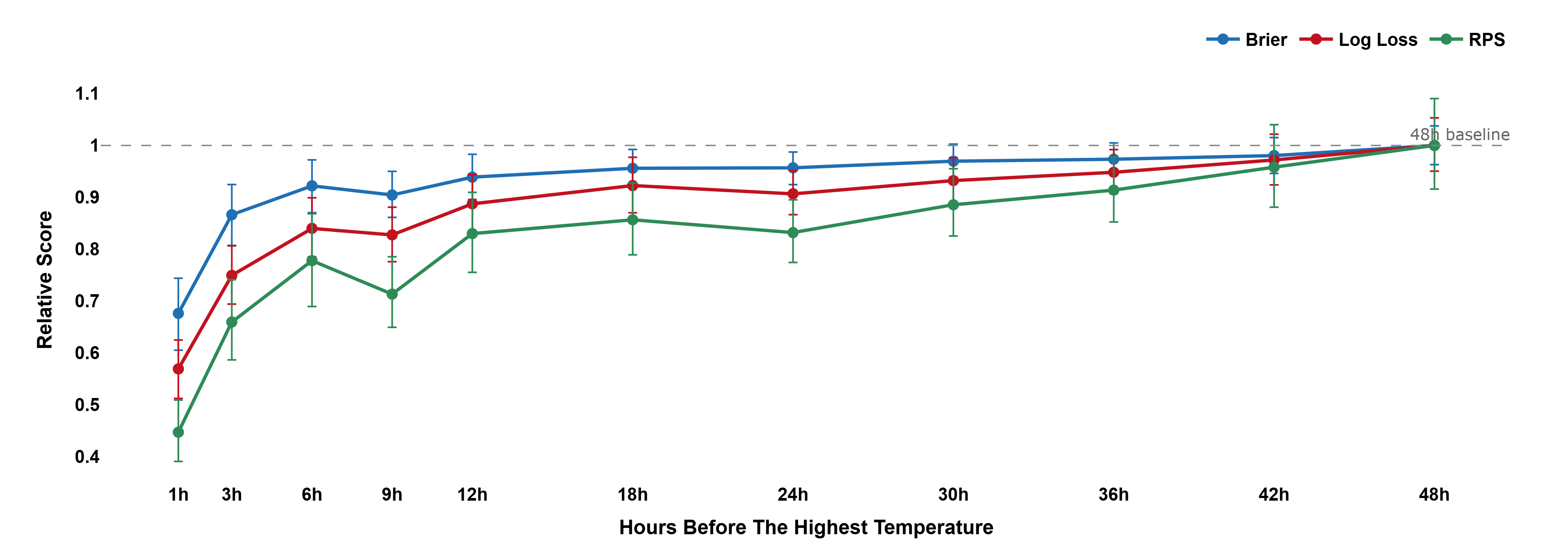}
\caption{Event-level Brier score, Log Loss, and RPS normalized to their 48-hour values. Error bars are 95\% confidence intervals from event-level bootstrap resampling. All three scores improve toward the daily temperature peak, with a sharper decline inside 12 hours; RPS shows the largest relative decrease.}
\label{fig:scores}
\end{figure}

Figure~\ref{fig:scores} uses the training sample, consistent with Figures~\ref{fig:reliability}--\ref{fig:heatmap}.

To compare their time paths on one scale, each score is divided by its value at 48 hours: $\RelativeScore_h=\Score_h/\Score_{48\mathrm{h}}$.
All three scores generally fall as the peak approaches (Fig.~\ref{fig:scores}). Improvement is gradual at 30--48 hours and becomes much faster inside 12 hours; each score is lowest at 1 hour. The faster late decline is consistent with updated forecasts and observations narrowing the plausible temperature range. RPS falls most, indicating that late prices concentrate not only on the winning bucket but also around nearby temperature buckets. The remaining favorite bias is therefore a local error within a distribution that is becoming more accurate overall.

\subsection{B-Spline Recalibration}

Reliability errors differ across low, middle, and high prices, with changes in both direction and slope across the three regions. A single intercept or slope cannot adjust these regions separately. A cubic B-spline gives a smooth nonlinear correction, while two internal knots provide separate local adjustments without allowing excessive changes in curvature. For $z=\logit(p)$,
\begin{equation}
\logit(\hat p)=w_0+\sum_{j=1}^{6}w_jB_j(z),
\label{eq:spline}
\end{equation}
where $B_j$ are cubic B-spline basis functions and $w_0,\ldots,w_6$ are fitted on the training data. The knots are the 0.35 and 0.65 quantiles of training logit prices, giving separate adjustment space to the three price regions. Corrected probabilities are normalized before scoring.

\begin{table}[t]
\caption{Event-level scores for all ten cities in the July test data. Lower is better. $\Delta$ is Raw minus Calibrated; 95\% confidence intervals are for the score differences.}
\label{tab:calibration}
\centering
\footnotesize
\setlength{\tabcolsep}{3pt}
\begin{tabular}{@{}lrrrc@{}}
\toprule
Metric & Raw & Calibrated & $\Delta$ & $\Delta$ 95\% CI \\
\midrule
Brier & 0.7045 & 0.7044 & +0.00008 & $[-0.00052,\,+0.00069]$ \\
Log Loss & 1.4450 & 1.4445 & +0.00056 & $[-0.00408,\,+0.00483]$ \\
RPS & 0.5705 & 0.5700 & +0.00055 & $[-0.00019,\,+0.00127]$ \\
\bottomrule
\end{tabular}
\end{table}

Table~\ref{tab:calibration} gives results for all ten cities in July. Point estimates improve slightly, but all 95\% confidence intervals for $\Delta$ include zero, providing no clear evidence of improvement. Most middle- and low-price bins are already close to calibration, while the main error is localized in the smaller high-price subset. Adjusting that subset has limited effect on scores that aggregate every bucket in an event. Table~\ref{tab:calibration} reports mean event-level scores over the same events for raw and calibrated forecasts, without adjustment for differences in bucket count.

\subsection{Buy-NO Test}

A negative range bias means that YES is overpriced, so buying NO takes the opposite side of the error. The tests use the windows in which each error is strongest: 1--12 hours for favorites and 30--48 hours for longshots. The favorite rule selects YES prices from 0.75 to 0.95, and the longshot rule selects prices from 0 to 0.10. The Buy-NO tests use the July test sample, with price ranges and horizon windows fixed from the training analysis. Each trade uses \$10 notional. Fees are calculated as $F=Cf\,r(1-r)$, where $r=1-p$ is the NO price, $C=10/r$ is the number of shares, and $f=0.05$ is the fee rate. The tests diagnose the economic direction of mispricing using mean taker prices. The Buy-NO calculations exclude order-book depth, bid--ask spreads, and slippage, so the reported returns do not establish executable trading profits. Win rate is the proportion of purchased NO shares that win.

\begin{table}[t]
\caption{Buy-NO diagnostics in the two strongest overpricing windows. Each trade uses ten dollars of notional. The 95\% confidence intervals use event-level bootstrap resampling and are reported in cents per dollar of notional.}
\label{tab:trading}
\centering
\footnotesize
\begin{tabular}{@{}lrr@{}}
\toprule
Metric & Favorite & Longshot \\
Window & 1--12 h & 30--48 h \\
\midrule
Total notional & \$1,840 (184 trades) & \$119,240 (11,924 trades) \\
Pre-fee PnL / ROI & +\$1,011 / +54.96\% & $-$\$153 / $-$0.13\% \\
Post-fee PnL / ROI & +\$933 / +50.71\% & $-$\$276 / $-$0.23\% \\
Post-fee PnL per trade & +\$5.07 & $-$\$0.02 \\
NO-share win rate & 20.7\% & 97.8\% \\
95\% CI (c/\$1) & $[1.76,\,102.98]$ & $[-0.47,\,0.03]$ \\
\bottomrule
\end{tabular}
\end{table}

Table~\ref{tab:trading} shows a 20.7\% NO win rate for the favorite rule. NO is cheap in this range, so relatively few wins produce large payoffs that offset the losing trades and fees, yielding a 50.71\% post-fee ROI in this sample, with a wide confidence interval. The longshot rule has a 97.8\% NO win rate, but NO costs nearly one dollar: most wins yield small gains, while rare YES wins cause large losses. Its pre-fee ROI is already near zero ($-0.13\%$), and fees reduce it to $-0.23\%$; the confidence interval includes zero. This contrast shows why a high win rate need not imply profit and why the favorite bias has the larger economic effect in this sample.

\balance
\section{Discussion and Conclusion}

The favorite--longshot bias in weather markets is strongly time-asymmetric. Longshots are mildly overpriced at distant horizons, and the error fades as weather information arrives. Favorites become much more overpriced during the final 1--12 hours. One possible behavioral interpretation is a two-stage process. Under high uncertainty, attention is spread across many cheap outcomes, including unlikely temperature buckets. Near the peak, most unlikely buckets have been removed, but confidence in the remaining high-price bucket rises beyond its observed win rate. The main weather-market error is therefore concentrated at the favorite end. Trader-level analysis would be needed to test this interpretation.

Full-distribution accuracy still improves as the peak approaches. This result does not conflict with the local favorite error: most middle- and low-price buckets are near calibration, so a strong error in a smaller high-price group has limited influence on aggregate event scores. B-spline point estimates improve only slightly, with all $\Delta$ confidence intervals spanning zero. The Buy-NO contrast adds an economic view: the inexpensive NO side of an overpriced favorite can offset a low win rate, while the expensive NO side of a longshot leaves little gain after fees despite winning often. Joint analysis of price range, horizon, and payoff structure reveals behavior that an overall calibration average can hide.

The close event-wtd and station-eq curves further indicate that near-term favorite overpricing is shared across the sample rather than created by one heavily represented city.

The evidence is limited to daily maximum-temperature markets on Polymarket, ten cities, and a one-month test period. Event counts, liquidity, and weather-information quality also differ across cities. Longer samples and other weather contract types are needed to test whether near-term favorite overpricing remains stable. The present results assess market calibration and do not establish an advantage over public weather forecasts. Future work will compare market probabilities with public probabilistic weather forecasts using matched information times and temperature buckets.

\section*{Author Contributions}

Shouchen Qiu collected and cleaned the data, conducted all experiments, and analyzed and interpreted the results. Huaijun Qiu contributed to the research design and framework and drafted the manuscript.

\end{document}